%% file: NuPhys2015.tex
\documentclass[11pt]{article}
\usepackage{graphicx}
\usepackage{amsmath,amssymb}
\usepackage{caption}
\usepackage{subcaption}
\usepackage{xcolor} 
\usepackage{cite} 
\usepackage{feynmf}

\newcommand\pubnumber{IPPP/16/44\\NuPhys2015-Weiland}
\newcommand\pubdate{\today}

\def\IPPP{Institute for Particle Physics Phenomenology, Department
  of Physics, Durham University, South Road, Durham DH1 3LE,
  United~Kingdom}

\def\IFLP{IFLP, CONICET - Dpto. de F\'{\i}sica, Universidad Nacional de La Plata, \\
C.C. 67, 1900 La Plata, Argentina}
\def\IFT{Departamento de F\'{\i}sica Te\'orica and Instituto de F\'{\i}sica Te\'orica, IFT-UAM/CSIC,\\
Universidad Aut\'onoma de Madrid, Cantoblanco, 28049 Madrid, Spain}

\def\Title#1{\begin{center} {\Large #1 } \end{center}}
\def\Author#1{\begin{center}{ \sc #1} \end{center}}
\def\Address#1{\begin{center}{ \it #1} \end{center}}

\newcommand\pubblock{\rightline{\begin{tabular}{l} \pubnumber\\
         \pubdate  \end{tabular}}}
\newenvironment{Abstract}{\begin{quotation}  }{\end{quotation}}
\newenvironment{Presented}{\begin{quotation} \begin{center} 
             PRESENTED AT\end{center}\bigskip 
      \begin{center}\begin{large}}{\end{large}\end{center} \end{quotation}}
\def\Acknowledgements{\bigskip  \bigskip \begin{center} \begin{large}
             \bf ACKNOWLEDGEMENTS \end{large}\end{center}}

\input econfmacros.tex

\begin{document}
\begin{titlepage}
\pubblock

\vfill
\Title{Lepton Flavour Violating Higgs Decays in the (SUSY) Inverse Seesaw}
\vfill
\Author{E. Arganda$^{\dagger}$, M.J. Herrero$^\ddagger$, X. Marcano$^\ddagger$, C. Weiland$^{\S,}$\footnote{Speaker}, }
\Address{$^\dagger$\IFLP \\ $^\ddagger $\IFT \\ $^\S$\IPPP}
\vfill
\begin{Abstract}
The observation of charged lepton flavour violation would be a smoking gun for new physics and could help
in pinpointing the mechanism at the origin of neutrino masses and mixing. We present here our recent 
studies of lepton flavour violating Higgs decays in the inverse seesaw and its supersymmetric embedding, two examples of low-scale seesaw mechanisms. We predict branching ratios 
as large as $10^{-5}$ for the decays 
$h\rightarrow \tau \mu$ and $h \rightarrow \tau e$ in the inverse seesaw, which can be probed in future colliders. Supersymmetric contributions can enhance the branching ratio of 
$h\rightarrow \tau \mu$ up to $1\%$, making it large enough to explain the small excess observed by ATLAS and CMS.
\end{Abstract}
\vfill
\begin{Presented}
NuPhys2015, Prospects in Neutrino Physics\\
Barbican Centre, London, UK,  December 16--18, 2015
\end{Presented}
\vfill
\end{titlepage}
\def\thefootnote{\fnsymbol{footnote}}
\setcounter{footnote}{0}

\section{Introduction}

The observation of neutrino oscillations proves that neutrinos are massive particles that mix and
calls for an extension of the Standard Model. Many neutrino mass generating mechanisms have been proposed and the 
discovery of a Higgs boson in 2012~\cite{Aad:2012tfa,Chatrchyan:2012xdj} with a mass of $m_h = 125.09 \pm 0.21 \mathrm{(stat.)} \pm 0.11 \mathrm{(syst.)\,GeV}$~\cite{Aad:2015zhl} opens new avenues 
to probe these models. Since neutrino oscillations violate lepton flavour conservation, particularly well-motivated observables are charged lepton flavour violating (cLFV) processes. Both CMS
and ATLAS collaborations have searched for cLFV Higgs decays~\cite{CMS:2015udp,Khachatryan:2015kon,Aad:2016blu}
and both have observed excesses in the $h\rightarrow \tau \mu$ channel, which translates into $\mathrm{BR}=0.84^{+0.39}_{-0.37}\%$ for CMS~\cite{Khachatryan:2015kon} and 
$\mathrm{BR}=0.53\pm0.51\%$ for ATLAS~\cite{Aad:2016blu}. These results and future improvements in sensitivity call for a study of their impact on neutrino mass models. We focus
here on two low-scale seesaw mechanisms, the inverse seesaw and its supersymmetric (SUSY) realization.

\section{The (SUSY) inverse seesaw}

One of the simplest and most appealing extensions of the Standard Model (SM) that generates neutrino masses and mixing is the inverse
seesaw (ISS)~\cite{Mohapatra:1986aw,Mohapatra:1986bd,Bernabeu:1987gr} where the SM is extended by adding three pairs of fermionic singlets with opposite
lepton number, denoted here by $\nu_{Ri}$ and $X_j$, with $i,j=1,2,3$. The following neutrino Yukawa interactions and mass terms are thus added to the SM Lagrangian~\cite{Arganda:2014dta}:
\begin{equation}
 \label{LagrangianISS}
 \mathcal{L}_\mathrm{ISS} = - Y^{ij}_\nu \overline{L_{i}} \widetilde{H} \nu_{Rj} - M_R^{ij} \overline{\nu_{Ri}^C} X_j - \frac{1}{2} \mu_{X}^{ij} \overline{X_{i}^C} X_{j} + h.c.\,,
\end{equation}
with $L$ the SM lepton doublet, $H$ the SM Higgs doublet, $\widetilde{H}=\imath \sigma_2 H^*$, $Y_\nu$ the $3\times3$ neutrino Yukawa coupling matrix, $M_R$ is a lepton number
conserving $3\times3$ mass matrix, and $\mu_X$ a Majorana $3\times3$ symmetric mass matrix. Since the latter controls the size of the lepton number violation, its smallness is 
natural~\cite{'tHooft:1979bh}. After electroweak symmetry breaking, the light neutrino mass matrix is given by~\cite{GonzalezGarcia:1988rw}
\begin{equation} \label{Mlight}
M_{\mathrm{light}} \simeq m_D {M_R^T}^{-1} \mu_X M_R^{-1} m_D^T\,,
\end{equation}
where $m_D=Y_\nu \langle H\rangle$, while heavy neutrinos form pseudo-Dirac pairs whose mass is approximately given by the eigenvalues of $M_R$ with the splitting within a pair 
controlled by $\mu_X$. Since the light neutrino masses
are suppressed by $\mu_X$, the ISS can naturally accommodate light neutrinos at the eV scale with $Y_\nu\sim\mathrm{O}(1)$ and a seesaw scale around the electroweak scale. 

The SUSY ISS is the simplest supersymmetric embedding of the ISS. 
It is defined by the superpotential:
\begin{equation}
 W=W_\mathrm{MSSM} + \varepsilon_{ab} \widehat N Y_\nu \widehat H^b_2 \widehat L^a + \widehat N M_R \widehat X + \frac{1}{2} \widehat X \mu_X \widehat X\,,
\end{equation}
with $\varepsilon_{12}=1$, $\widehat H_1$ and $\widehat H_2$ the down-type and up-type Higgs bosons.
The reader can find the corresponding soft SUSY breaking Lagrangian in our main article~\cite{Arganda:2015naa}. Importantly, all soft SUSY breaking masses are taken to be flavour diagonal,
making sure that the only source of cLFV is the neutrino Yukawa coupling $Y_\nu$.

\section{cLFV Higgs decays in the inverse seesaw}
\label{cLFVHDiss}

In this work, we have considered the full set of one-loop diagrams contributing to the cLFV Higgs decay rates in the ISS. All diagrams, the details of the calculation as well as the details of the 
constraints implementation can be found in~\cite{Arganda:2014dta}. 
In particular, we make use of two parametrizations in order to reproduce light neutrino masses and mixing in agreement with oscillation data~\cite{GonzalezGarcia:2012sz}, a
Casas-Ibarra~\cite{Casas:2001sr} parametrization for $Y_\nu$, modified for the ISS, and a new parametrization for $\mu_X$:
\begin{equation}
\mu_X=M_R^T ~m_D^{-1}~ U_{\rm PMNS}^* m_\nu U_{\rm PMNS}^\dagger~ {m_D^T}^{-1} M_R\,,
\end{equation}
where $U_{\rm PMNS}$ is the $3\times3$ Pontecorvo-Maki-Nakagawa-Sakata matrix for neutrino mixing.

\begin{figure}[t!]
\begin{center}
\includegraphics[width=0.43\textwidth]{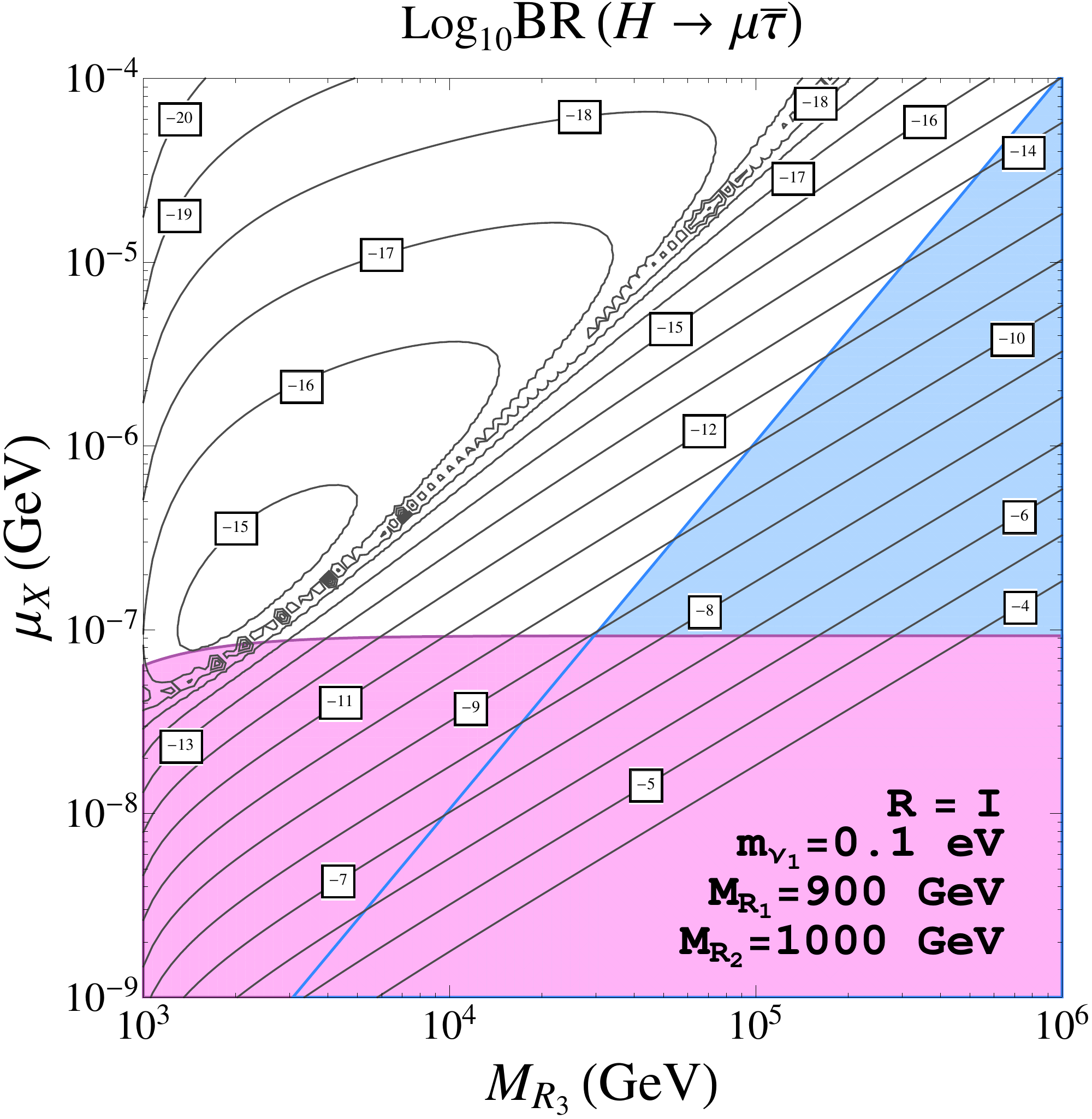}
\includegraphics[width=0.56\textwidth]{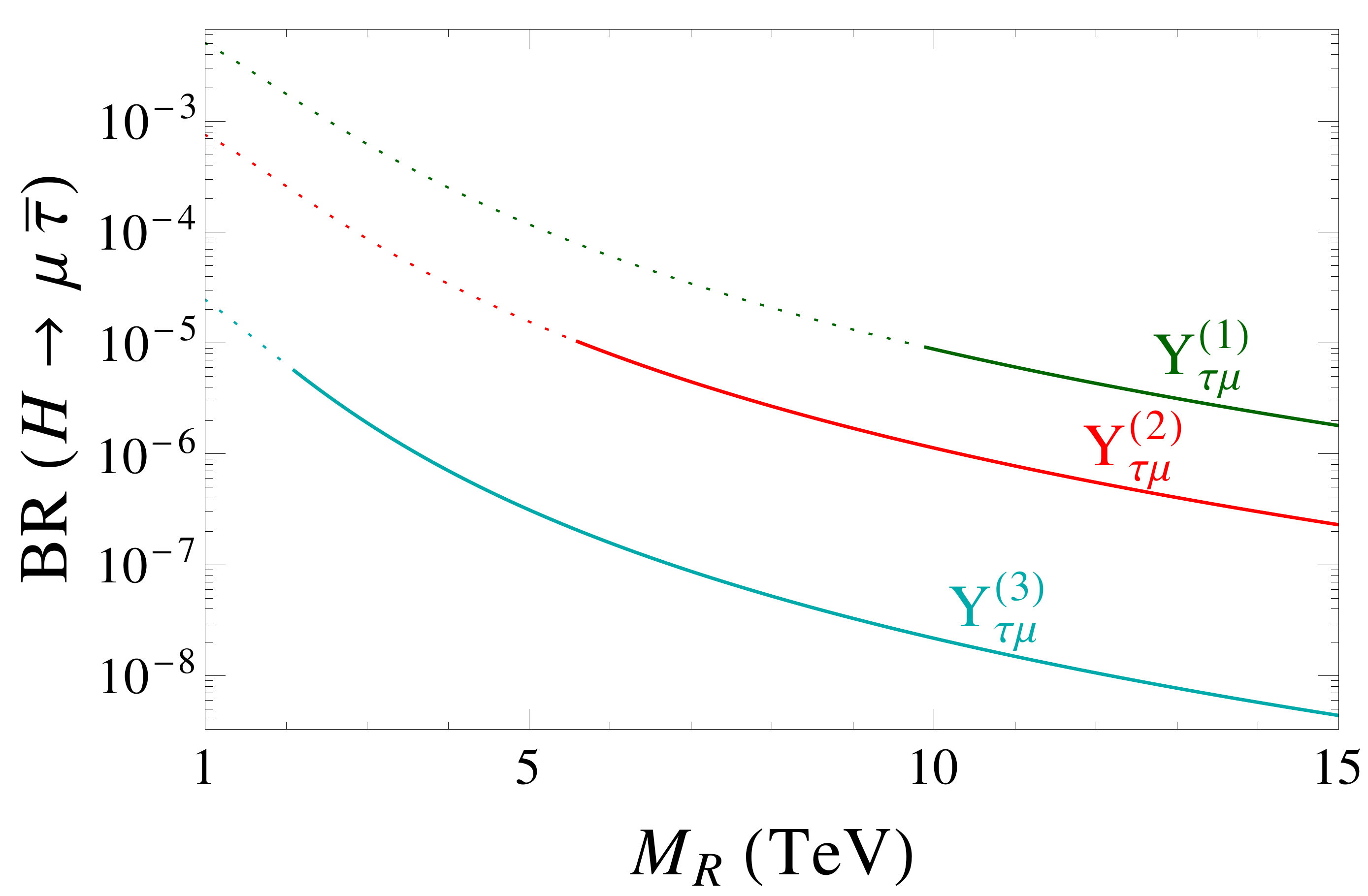}
\caption{Left panel: Contour lines for BR$(h \to \mu \bar \tau)$ in the $(M_{R_3},\mu_X)$ plane using the modified Casas-Ibarra parametrization.
The pink area is excluded by BR$(\mu \to e \gamma)$ and the blue area is exclude by the non-perturbative $Y_\nu$. Right panel: BR$(h \to \mu \bar \tau)$ as a function of $M_R$ using the
$\mu_X$-parametrization with $f=\sqrt{6\pi}$. Dotted lines
indicate excluded input values leading to BR$(\tau \to \mu \gamma)$ above the present experimental bound.}
\label{maxcLFVHD}
\end{center}
\end{figure}

The left panel of fig.~\ref{maxcLFVHD} shows BR$(h \to \mu \bar \tau)$ as a function of the two most relevant parameters in the case of hierarchical heavy neutrino. Since we use a
modified Casas-Ibarra parametrization here, $Y_\nu$ grows with $M_R$, which leads to larger $\mathrm{BR}(h \rightarrow \mu \bar \tau)$ at a larger $M_{R_3}$. In contrast, $Y_\nu$
decreases when $\mu_X$ increases, meaning that $\mathrm{BR}(h \rightarrow \mu \bar \tau)$ grows when $\mu_X$ decreases. Circles and lines correspond to two regions with different dependences
on the seesaw parameters, which is specific to cLFV Higgs decays. Radiative cLFV decays only exhibit lines for example. We can clearly see that, in this case,
$\mathrm{BR^{max}}(h\rightarrow \mu \bar{\tau}) \sim10^{-9}$ due to the stringent experimental upper limit on BR$(\mu \to e \gamma)$~\cite{Adam:2013mnn}.

The right panel of fig.~\ref{maxcLFVHD} displays BR$(h \to \mu \bar \tau)$ as a function of the seesaw scale in the $\mu_X$-parametrization with degenerate heavy neutrinos. We are now free to
choose as input Yukawa couplings with $(Y_\nu Y_\nu^\dagger)_{12}=0$, which suppresses $\mathrm{BR}(\mu \rightarrow e\, \gamma)$. Three examples are
\begin{equation}
Y_{\tau \mu}^{(1)}=f\left(\begin{array}{ccc}
0&1&-1\\0.9&1&1\\1&1&1
\end{array}\right)~,~
Y_{\tau \mu}^{(2)}=f\left(\begin{array}{ccc}
0&1&1\\1&1&-1\\-1&1&-1
\end{array}\right)~,~
Y_{\tau \mu}^{(3)}=f\left(\begin{array}{ccc}
0&-1&1\\-1&1&1\\0.8&0.5&0.5
\end{array}\right)\,, \nonumber
\label{Ytmmax}
\end{equation}
where the constraint $\frac{|Y^{ij}_\nu|^2}{4\pi} \leq 1.5$ translates into $f\leq\sqrt{6\pi}$. The main constraint in this case is $\mathrm{BR}(\tau\rightarrow \mu \gamma)$~\cite{Aubert:2009ag},
which decreases faster
than $\mathrm{BR}(h\rightarrow \mu \bar{\tau})$. As a consequence, we find that $\mathrm{BR^{max}} (h \rightarrow \mu \bar \tau)\sim 10^{-5}$, a result that remains valid with
hierarchical heavy neutrinos. Similarly, we found $\mathrm{BR^{max}} (h \rightarrow e\, \bar \tau)\sim 10^{-5}$. Details of the analysis, including the dependence on other parameters,
can be found in~\cite{Arganda:2014dta}. Branching ratios of this size, while they cannot explain the small LHC excess, can be probed at future colliders.

\section{cLFV Higgs decays in the SUSY inverse seesaw}

For this study, we have calculated the full one-loop SUSY contributions in the mass basis to the SM-like Higgs boson, denoted by h and taken to be the lightest CP-even Higgs boson.
The contributing diagrams as well as the complete analytical expressions can be found
in~\cite{Arganda:2015naa}.  We have taken into account various experimental constraints like low energy neutrino data and cLFV radiative decays, choosing as examples two benchmark points
with a Higgs boson mass within 1$\sigma$ of the central value of the latest CMS and ATLAS combination and with supersymmetric spectra allowed by ATLAS and CMS
searches.

\begin{figure}[t!]
\begin{center}
\includegraphics[width=0.49\textwidth]{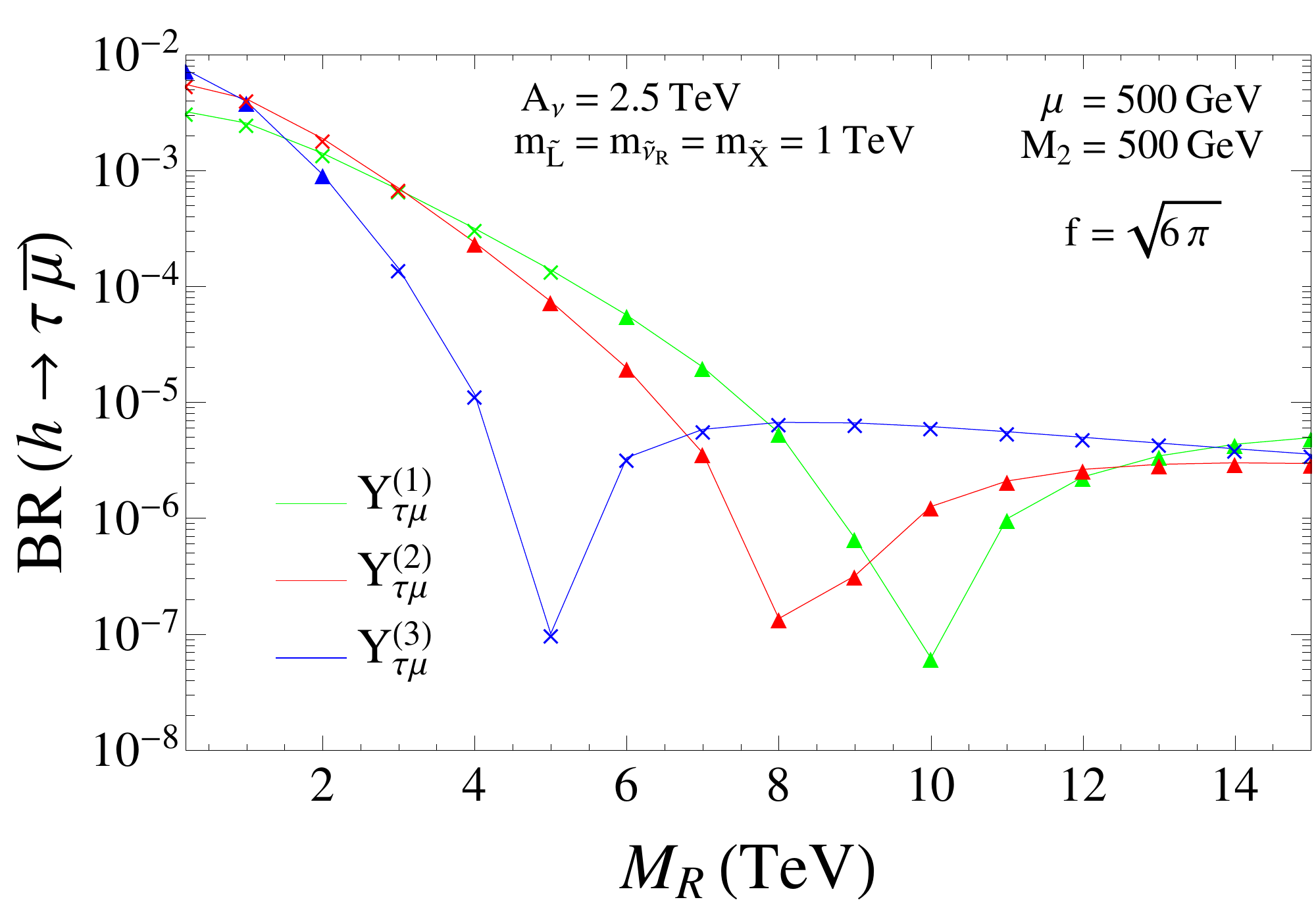}
\includegraphics[width=0.49\textwidth]{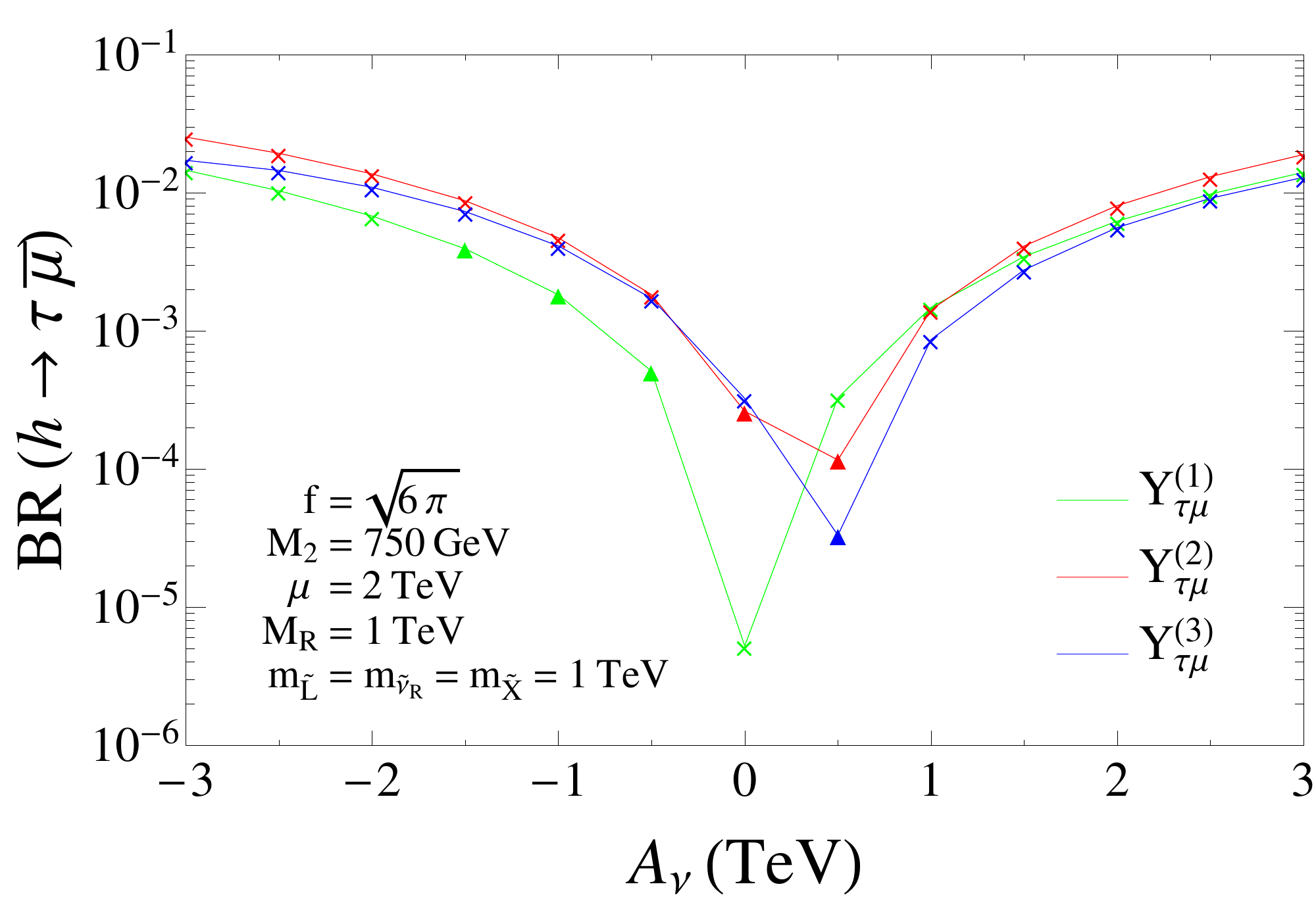}
\caption{BR($h \to \tau \bar \mu$) in the SUSY-ISS using the $\mu_X$-parametrization, $m_A =$ 800 GeV and $M_0=$ 1 TeV. Left panel:  BR($h \to \tau \bar \mu$) as a
function of $M_R$ with $\tan\beta = 10$.
Right panel: Dependence of BR($h \to \tau \bar \mu$) on $A_\nu$ with $\tan\beta =$ 5. Crosses (triangles) represent points
in the SUSY-ISS parameter space excluded (allowed) by the $\tau \to \mu \gamma$ upper limit~\cite{Aubert:2009ag}.}
\label{cLFVHDsusy}
\end{center}
\end{figure}

We present in fig.~\ref{cLFVHDsusy} the predictions of BR($h \to \tau \bar \mu$) as a function of the seesaw scale and the sneutrino trilinear coupling for the
three neutrino Yukawa textures presented in the previous section~\ref{cLFVHDiss}, suppressing cLFV in the $\mu-e$ sector. The most stringent constraint is thus the related cLFV
radiative decay $\tau \to \mu \gamma$. On the left panel, two different behaviours can be observed. At low $M_R$, the dominant contribution comes from sneutrino-chargino loops while 
slepton-neutralino loops dominate at large $M_R$. This comes from the roughly linear right-handed sneutrino mass dependence on $M_R$, leading to their decoupling at large $M_R$. On the right panel,
we can see that large values of $A_\nu$ increases both cLFV radiative and Higgs decays when the sneutrino-chargino contributions dominate. However, the exact value of $A_\nu$ where they reach their
minimum can be different, leading to an enhanced BR($h \to \tau \bar \mu$) with BR($\tau \to \mu \gamma$) in agreement with experimental limits. This leads to 
$\mathrm{BR^{max}}(h\rightarrow \tau \bar \mu)\sim1\%$, a value large enough to explain the CMS and ATLAS excesses. A more detailed discussion of the dependence on relevant parameters can be found
in~\cite{Arganda:2015naa}.

\section{Conclusion}

Following the discovery of a Higgs boson at the LHC, searches for cLFV Higgs decays offer a new way to probe the mechanism at the origin of neutrino masses and mixing. We have presented here 
the results of two studies of cLFV Higgs decays in the inverse seesaw and its supersymmetric realization, showing first that they are complementary to cLFV radiative decays because of their different
dependence on the seesaw parameters. In the non-supersymmetric inverse seesaw, $\mathrm{BR}(h \rightarrow \mu \bar \tau)$ and $\mathrm{BR}(h \rightarrow e \bar \tau)$ as large as $10^{-5}$ can
be expected, which can be probed by future colliders. Supersymmetric contributions to $h\rightarrow \tau \mu$ can lead to branching ratios up to $1\%$, which could explain the CMS and ATLAS excesses. If the seesaw scale is low enough to generate these large branching ratios, the scenarios considered here could lead to a substantial production of heavy neutrinos at
the LHC, leading to specific final states like $\mu \tau jj$ with $M_{jj}=M_W$~\cite{Arganda:2015ija}.

\Acknowledgements
C.~W. received financial support as an International Research Fellow of the Japan Society for the Promotion of Science and
from the European Research Council under the European Union's Seventh Framework Programme (FP/2007-2013) / ERC Grant NuMass agreement n. [617143] during different stages of this work.
E.~A. acknowledges the support of ANPCyT, Argentina. X. M. is supported through the FPU grant AP-2012-6708.
Some of the work presented here was partially supported by the European Union Grant No. FP7 ITN
INVISIBLES (Marie Curie Actions, Grant No. PITN- GA-2011- 289442), by the European Union Horizon 2020 research and innovation programme under the Marie Skłodowska-Curie grant agreements
No 690575 and
No 674896,
by the CICYT through Grant No. FPA2012-31880,  
by the Spanish Consolider-Ingenio 2010 Programme CPAN (Grant No. CSD2007-00042), 
and by the Spanish MINECO's ``Centro de Excelencia Severo Ochoa'' Programme under Grant No. SEV-2012-0249.

\end{document}

%% file: econfmacros.tex
\def\beq{\begin{equation}}
\def\eeq#1{\label{#1}\end{equation}}
\def\eeqn{\end{equation}}

\def\beqa{\begin{eqnarray}}
\def\eeqa#1{\label{#1}\end{eqnarray}}
\def\eeqan{\end{eqnarray}}

\let\bar=\overbar

\def\Dslash{\not{\hbox{\kern-4pt $D$}}}
\def\dslash{\not{\hbox{\kern-2pt $\del$}}}

\def\msb{{\bar{\ssstyle M \kern -1pt S}}}

